\documentclass[12pt,a4paper]{article}
\pdfoutput=1
\usepackage{graphicx}
\usepackage[T1]{fontenc}
\usepackage[sc,osf]{mathpazo}
\usepackage{a4wide}  
\usepackage{latexsym,amsthm,amsfonts,amsmath,mathrsfs,amssymb}
\usepackage{booktabs} 
\newcommand{\req}[1]{(\ref{#1})} 
\usepackage[unicode,implicit]{hyperref}
\hypersetup{%
  pdftitle    = {Non-Abelian black holes in string theory}
  pdfkeywords = {Yang-Mills, instanton, black hole, string theory, entropy,
    branes, supergravity, supersymmetry},
  pdfauthor   = {Pablo A. Cano, Patrick Meessen, Tomas Ortin, Pedro F. Ramirez},
  plainpages  = true,
  colorlinks  = true,
  citecolor   = blue,
  urlcolor    = red,
  linkcolor   = black
}
\newcommand{\hepth}[1]{{\tt
\href{http://www.arXiv.org/abs/hep-th/#1}{hep-th/#1}}}

\newcommand{\arxiv}[1]{{\tt arXiv:\href{http://www.arXiv.org/abs/#1}{#1}}}
\newcommand{\doi}[1]{{\tt DOI:\href{http://dx.doi.org/#1}{#1}}}

\usepackage{tikz}\newcommand{\FPAUO}[2]{
\tikz[scale=.13,
         Uniovi/.style={color=green!51!blue, fill=green!51!blue}
 ] {
 \fill[Uniovi] (0,0) circle (10);
 \fill[white] (0,7) circle (1.5);
 \draw[Uniovi] (-2,7.5) rectangle (2,5.5);
 \fill[white] (-0.3,6.6) rectangle (0.3,0);   
 \fill[white] ( -0.9,6.2) rectangle (.9 ,5.6);
 \fill[white] (-1.4, 5.2) rectangle (1.4, 4.6);
 \fill[white] (0,0) ellipse (3.5 and 4);
 \fill[Uniovi] (-2.5,0.3) rectangle (2.5,-0.3);
 \fill[Uniovi] (-2,2.3) rectangle (2,1.7);
 \fill[Uniovi] (-2,-2.3) rectangle (2,-1.7);
 \fill[white] (-4.5,5.5) rectangle (-2.7,4.9);
 \fill[white] (-3.9,6.1) rectangle (-3.3,4.3);
 \fill[white] (4.5,5.5) rectangle (2.7,4.9);
 \fill[white] (3.9,6.1) rectangle (3.3,4.3);
 \foreach \x in { 0,..., 3 }
   \foreach \y in { 0,...,\x}
    {
     \fill[white] (-6-\x*0.7+\y*1.4,3.5-\x *1.97) -- (-5.6-\x*0.7+\y*1.4,2.4-\x *1.97) -- (-6.4-\x*0.7+\y*1.4,2.4-\x *1.97) -- cycle;
     \fill[white] (6-\x*0.7+\y*1.4,3.5-\x *1.97) -- (5.6-\x*0.7+\y*1.4,2.4-\x *1.97) -- (6.4-\x*0.7+\y*1.4,2.4-\x *1.97) -- cycle;
   };
 \draw (0,-6) node[
                               text centered, 
                               color=white, 
                               font={\fontsize{8}{4}\sffamily\selectfont}
                             ] {FPAUO-#1/#2};
}} 
\usepackage{pgfplots}    
\makeatletter
\@addtoreset{equation}{section}
\makeatother

\begin{document}

\begin{flushright}
\small
\FPAUO{17}{05}\\
IFT-UAM/CSIC-17-025\\
\texttt{arXiv:1704.01134 [hep-th]}\\
December 14\textsuperscript{th}, 2017 
\normalsize
\end{flushright}

\vspace{0cm}

\begin{center}

{\Large {\bf {Non-Abelian black holes in string theory}}}
 
\vspace{1.5cm}

\renewcommand{\thefootnote}{\alph{footnote}}
{\sl\large Pablo A.~Cano$^{1}$}${}^{,}$\footnote{E-mail: {\tt pablo.cano [at] uam.es}},
{\sl\large Patrick Meessen$^{2}$}${}^{,}$\footnote{E-mail: {\tt meessenpatrick [at] uniovi.es}},
{\sl\large Tom\'{a}s Ort\'{\i}n$^{1}$}${}^{,}$\footnote{E-mail: {\tt Tomas.Ortin [at] csic.es}}
{\sl\large and Pedro F.~Ram\'{\i}rez$^{1}$}${}^{,}$\footnote{E-mail: {\tt p.f.ramirez [at]  csic.es}},

\setcounter{footnote}{0}
\renewcommand{\thefootnote}{\arabic{footnote}}

\vspace{1cm}

${}^{1}${\it Instituto de F\'{\i}sica Te\'orica UAM/CSIC\\
C/ Nicol\'as Cabrera, 13--15,  C.U.~Cantoblanco, E-28049 Madrid, Spain}\\ 
\vspace{0.3cm}

${}^{2}${\it HEP Theory Group, Departamento de F\'{\i}sica, Universidad de Oviedo\\
  Avda.~Calvo Sotelo s/n, E-33007 Oviedo, Spain}\\

\vspace{.2cm}


{\bf Abstract}

\end{center}

\begin{quotation}
  {\small We study a family of 5-dimensional non-Abelian black holes that can
    be obtained by adding an instanton field to the well-known D1D5W Abelian
    black holes. Naively, the non-Abelian fields seem to contribute to the
    black-hole entropy but not to the mass due to their rapid fall-off at
    spatial infinity. By uplifting the 5-dimensional supergravity solution to
    10-dimensional Heterotic Supergravity first and then dualizing it into a
    Type-I Supergravity solution, we show that the non-Abelian fields are
    associated to D5-branes dissolved into the D9-branes (dual to the
    Heterotic ``gauge 5-branes'') and that their associated RR charge does
    not, in fact, contribute to the entropy, which only depends on the number
    of D-strings and D5 branes and the momentum along the D-strings, as in the
    Abelian case. These ``dissolved'' or ``gauge'' D5-branes do contribute to
    the mass in the expected form. The correct interpretation of the
    5-dimensional charges in terms of the string-theory objects solves the
    non-Abelian hair puzzle, allowing for the microscopic accounting of the
    entropy. We discuss the validity of the solution when $\alpha'$
    corrections are taken into account.}
\end{quotation}

\newpage
\pagestyle{plain}



\section*{Introduction}

One of the common features of black holes or black rings with genuinely
non-Abelian fields\footnote{That is: non-Abelian fields that cannot be related
  to an Abelian embedding via a (possibly singular) gauge transformation
  \cite{Smoller:1991ag}. Gauge transformations, whether regular or singular,
  have no effect whatsoever on the spacetime metric and, therefore, if the
  non-Abelian fields can be related to an Abelian embedding, the metric is
  effectively that of a solution with an Abelian field. This was the only kind
  of regular solutions thought to exist in the Einstein-Yang-Mills theory,
  basically because the non-Abelian fields were expected to behave at infinity
  like the Abelian ones \cite{Galtsov:1989ip,Ershov:1991nv,Bizon:1992pi}. See
  also See Refs.~\cite{Volkov:1998cc,Galtsov:2001myk} and references therein.}  in
Einstein-Yang-Mills (EYM) theory, where they are only known numerically
\cite{Volkov:1989fi,Bizon:1990sr}, or in $\mathcal{N}=2,d=4,5$ Super-EYM
(SEYM) theories
\cite{Meessen:2008kb,Meessen:2015nla,Meessen:2015enl,Ortin:2016bnl}, where
they are known analytically, is that their non-Abelian fields fall off at
spatial infinity so fast that they cannot be characterized by a conserved
charge. For this reason they are sometimes called ``colored'' black holes, as
opposed to ``charged'' black holes.  As a consequence, the parameters that
characterize the black holes must be understood as pure non-Abelian
hair.

In the SEYM case it has also been observed that the non-Abelian fields seem to
contribute in a non-trivial way to the BH entropy because their near-horizon
behavior is similar to that of their Abelian counterparts
\cite{Meessen:2008kb,Meessen:2015nla,Meessen:2015enl,Ortin:2016bnl}. Thus,
apparently, the entropy of these non-Abelian black holes and rings depends on
non-Abelian hair! If the BH entropy admits a microscopic interpretation, this
conclusion is clearly unacceptable.

In this paper we are going to solve this puzzle for a family of particularly
simple non-Abelian 5-dimensional black holes that can be embedded in String
Theory \cite{Meessen:2015enl} and which can be seen as the well-known 3-charge
D1D5W black-hole solutions discussed in
Ref.~\cite{Callan:1996dv}\footnote{More information on these black holes and
  the String Theory computation of their BH entropy can be found in
  Ref.~\cite{David:2002wn} and references therein.} with the addition of a
BPST instanton \cite{Belavin:1975fg}, which is genuinely non-Abelian in the
sense discussed above.\footnote{Technically, this family of black holes is a
  solution of the SU$(2)$-gauged ST$[2,6]$ model of $\mathcal{N}=1,d=5$
  supergravity. This model and the solution-generating technique used to
  obtain the black-hole family is described in full detail in an Appendix of
  Ref.~\cite{Cano:2017sqy}.}  The embedding is realized via Heterotic
Supergravity (that is: $\mathcal{N}=1,d=10$ supergravity coupled to vector
supermultiplets) without the terms of higher order in the curvature of the
torsionful spin connection which corresponds to the low-energy effective field
theory of the Heterotic Superstring. Our solution is an exact supergravity
solution but, clearly, the issue of $\alpha'$ corrections needs to be
addressed. As we show in Appendix~\ref{app-alpha}, the supergravity solution
we are studying is also good to order $\alpha'$ in Heterotic Superstring
theory, but only in the near-horizon region and needs to be
$\alpha'$-corrected elsewhere. Finding these corrections is a problem that we
will tackle in a forthcoming publication \cite{kn:CMOR} and, in the meantime,
one can work with the supergravity solution within the limits we just
mentioned. In particular, the supergravity solution should be enough to
characterize the different branes the black hole is ``made of''.

Back to the non-Abelian hair puzzle, in this case at least, the solution lies
in the correct interpretation of the different charges that characterize the
black hole. As we have shown in Ref.~\cite{Cano:2017sqy}, the charges that
count the underlying String Theory objects are combinations of the naive
ones. The correctly identified charges can be switched off one by one and,
switching off those that count the objects that give rise to the Abelian
charges (that is, setting to zero the number of D1s, D5s and the momentum) one
is left with the object that produces the net non-Abelian field. In 5
dimensions, this object is a globally regular, horizonless gravitating
instanton \cite{Cano:2017sqy} which, when uplifted to 10-dimensional Heterotic
Supergravity (the effective field theory of the Heterotic Superstring), is
nothing but Strominger's gauge 5-brane \cite{Strominger:1990et}.\footnote{For
  recent work on Abelian black-hole solutions of Heterotic Supergravity (with
  $R^{2}$terms, the Hull-Strominger system) see Ref.~\cite{Halmagyi:2016pqu}
  and references therein.}  In terms of these charges, as we will see, there
is a non-Abelian contribution to the mass and the non-Abelian contribution to
the entropy disappears, solving the puzzle.

This is a very important clue that we are going to apply to these
solutions. In Section~\ref{sec-solutions} we are going to introduce them and
rewrite them in terms of the charges that describe the underlying
String-Theory objects. In Section~\ref{sec-heterotic} we are going to uplift
them to 10-dimensional Heterotic Supergravity, a theory that has non-Abelian
vector fields in 10 dimensions, and, in Section~\ref{sec-STinterpretation} we
will reinterpret the solution in terms of intersections of fundamental
strings, solitonic 5-branes and gauge 5-branes, plus momentum along the
strings, and we will dualize it into a solution of Type-I Supergravity (the
effective field theory of Type-I Superstring Theory)
\cite{Dabholkar:1995ep,Hull:1995nu,Polchinski:1995df} with D-strings,
momentum, D5-branes and ``gauge D5-branes'', the duals of the gauge 5-branes,
also referred to as D5-branes dissolved into the D9 branes. Then, in
Section~\ref{sec-discussion} we discuss how this brane configuration leads to
the same entropy as the Abelian one, pointing to directions for future work.
Finally, in Appendix~\ref{app-alpha} we discuss the validity of our solution
of 10-dimensional Heterotic Supergravity as a solution of the Heterotic
Superstring taking into account $\alpha'$ corrections.

\section{5-dimensional non-Abelian black holes}
\label{sec-solutions}

We consider the SU$(2)$-gauged ST$[2,6]$ model of $\mathcal{N}=1,d=5$
supergravity, which can be obtained from $d=10$ Heterotic Supergravity by
compactification on $T^{5}$ followed by a truncation. This is most conveniently
done in two stages: first, compactification on $T^{4}$ followed by a
truncation to $\mathcal{N}=(2,0),d=6$ supergravity coupled to a tensor
multiplet and a triplet of SU$(2)$ vectors and, second, further
compactification on $S^{1}$. The first stage is almost trivial: all the
6-dimensional fields are identical (up to rescalings) to the first 6
components of the 10-dimensional ones. The second stage is described in detail
in Ref.~\cite{Cano:2016rls}.

This model is determined by the symmetric tensor
$C_{0xy}=\frac{1}{6}\eta_{xy}$, with $x,y=1,2,A$, $A,B,\ldots=3,4,5$ and
$\eta_{xy}=(+,-,-,-,-)$.\footnote{A more detailed description of this model
  can be found in Appendix~A of Ref.~\cite{Cano:2017sqy}, for instance.} The
$A,B,\ldots$ are adjoint SU$(2)$ indices.  The bosonic content of this model
consists of the metric $g_{\mu\nu}$, 3 Abelian vectors, $A^{0}$, $A^{1}$ and
$A^{2}$ a triplet of SU$(2)$ vectors $A^{A}$, and 5 scalars which we choose as
$\phi, k$ and $\ell^{A}$ where $\phi$ can be directly identified with the
10-dimensional heterotic dilaton and $k$ is the Kaluza-Klein scalar of the
last compactification from $d=6$ to $d=5$.

A particularly simple family of non-Abelian black-hole solutions of
$\mathcal{N}=1,d=5$ supergravity can be constructed by adding a BPST instanton
to the standard 3-charge solution
\cite{Bueno:2015wva,Meessen:2015enl,Cano:2017sqy}. The family of solutions is
determined by 3 harmonic functions $L_{0,\pm}$ which depend on three constants
$B_{0,\pm}$ satisfying $\frac{27}{2}B_{0}B_{+}B_{-}= 1$ and three independent
charges $q_{0,\pm}$

\begin{equation}
L_{0,\pm} = B_{0,\pm} + q_{0,\pm}/\rho^{2}\, ,
\end{equation}

\noindent
and a non-Abelian contribution that depends on the 5-dimensional gauge
coupling constant $g$ and on the instanton scale $\kappa$

\begin{equation}
\Phi^{2} 
\equiv 
\frac{2 \kappa^{4}}{3g^{2}\rho^{4}(\rho^{2} +\kappa^{2})^{2}}\, .   
\end{equation}

\noindent
The non-Abelian contribution appears combined with the harmonic function
$L_{0}$ as follows:

\begin{equation}
\tilde{L}_{0} \equiv  L_{0}- \tfrac{1}{3}\rho^{2}\Phi^{2}\, ,
\end{equation}

\noindent
and, since it goes like $1/\rho^{6}$ at spatial infinity while $L_{0}$ goes
like $B_{0}+q_{0}/\rho^{2}$, it is not expected to contribute to the
mass. However, both the Abelian and non-Abelian contributions diverge like
$1/\rho^{2}$ near the horizon at $\rho=0$, and, naively, one expects both of
them to contribute to the entropy. This can be manifest by rewriting $\tilde{L}_{0}$ as 

\begin{equation}
\tilde{L}_{0}
= 
B_{0}
+
(q_{0}- \frac{2}{9g^{2}})\frac{1}{\rho^{2}}
+
\frac{2}{9g^{2}}
\frac{\rho^{2} +2\kappa^{2}}{(\rho^{2} +\kappa^{2})^{2}}
\, ,
\end{equation}

\noindent
where we have combined Abelian and non-Abelian $1/\rho^{2}$ terms in
$\tilde{L}_{0}$, leaving a purely non-Abelian contribution which is finite at
$\rho=0$. As in Ref.~\cite{Cano:2017sqy}, we will call $\tilde{q}_{0}\equiv q_{0}-
\frac{2}{9g^{2}}$ the coefficient of the $1/\rho^{2}$ term.

The constants $B_{0,\pm}$ are related to the moduli \textit{i.e.}~the values of
the 2 scalars at infinity\footnote{We will relate the charges to the numbers
  of branes in $d=10$ after embedding the solution in Heterotic Supergravity.}
as follows

\begin{equation}
B_{0} = \tfrac{1}{3} e^{\phi_{\infty}}k_{\infty}^{-2/3}\, ,
\hspace{1cm}
B_{-} = \tfrac{2}{3} e^{-\phi_{\infty}}k_{\infty}^{-2/3}\, ,
\hspace{1cm}
B_{+} = \tfrac{1}{3} k_{\infty}^{4/3}\, .
\end{equation}

Is is convenient to use the functions $\tilde{\mathcal{Z}}_{0}\equiv
\tilde{L}_{0}/B_{0}$ and $\mathcal{Z}_{\pm}\equiv L_{\pm}/B_{\pm}$ and the
charges $\tilde{\mathcal{Q}}_{0}\equiv \tilde{q}_{0}/B_{0}=(q_{0}-
\frac{2}{9g^{2}})/B_{0}$ and $\mathcal{Q}_{\pm}\equiv q_{\pm}/B_{\pm}$. 

It is also convenient to transform the BPST instanton field from the gauge
used in Refs.~\cite{Meessen:2015enl,Ortin:2016bnl} to one in which the
10-dimensional solution will be easier to recognize:\footnote{The reason why
  this gauge was not used in Refs.~\cite{Meessen:2015enl,Ortin:2016bnl} is
  that, in it, the gauge field cannot be consistently reduced following
  Kronheimer.}${}^{,}$\footnote{Our conventions for the SU$(2)$ gauge fields
  are slightly different from the ones used in
  Refs.~\cite{Meessen:2015enl,Ortin:2016bnl}. Here the generators satisfy the
  algebra $[T_{A},T_{B}]=+\epsilon_{ABC}T_{C}$, the left-invariant
  Maurer-Cartan 1-forms are defined by $v_{L}\equiv -U^{-1}dU$ and the
  right-invariant ones by $v_{R}\equiv -dUU^{-1}$.  The gauge field strength
  is defined by $F= dA + gA\wedge A$.}

\begin{equation}
A^{A}_{R}
=
\frac{1}{g}\frac{\kappa^{2}}{(\kappa^{2}+\rho^{2})}v^{A}_{R}
\,\,\,\,\,
\longrightarrow 
\,\,\,\,\,
A^{A}_{L}
=
-\frac{1}{g}\frac{\rho^{2}}{(\kappa^{2}+\rho^{2})}v^{A}_{L},
\end{equation}

\noindent
The vector field strength is, evidently, the same, but the
Chern-Simons term is not and this difference will also affect the
10-dimensional 2-form. 

After all these transformations, the active fields of the solutions
are\footnote{Since we are going to use hats to denote 10-dimensional fields, we
  have removed the hats that we use in our notation for the metric function
  $f$.}

\begin{equation}
\label{eq:3chargebh1}
\begin{array}{rclrcl}
ds^{2} 
& = & 
f^{2}dt^{2}-f^{-1}(d\rho^{2}+\rho^{2}d\Omega_{(3)}^{2})\, ,
& & & 
\\
& & & & & \\
A^{0} 
& = &
-\sqrt{3}e^{-\phi_{\infty}}k_{\infty}^{2/3}{\displaystyle\frac{dt}{\tilde{\mathcal{Z}}_{0}}}\, ,
\hspace{1cm}
&
A^{1}+A^{2} 
& = & 
-2\sqrt{3}k_{\infty}^{-4/3}{\displaystyle\frac{ dt}{\mathcal{Z}_{+}}}\, , 
\\
& & & & & \\
A^{A} 
& = &
{\displaystyle
-\frac{1}{g}\frac{\rho^{2}}{(\kappa^{2}+\rho^{2})}v^{A}_{L}\, ,  
}
&
A^{1}- A^{2} 
& = & 
-\sqrt{3} e^{\phi_{\infty}}k_{\infty}^{2/3}{\displaystyle\frac{dt}{\mathcal{Z}_{-}}}\, , 
\\
& & & & & \\
e^{2\phi}
& = &
e^{2\phi_{\infty}}{\displaystyle\frac{\tilde{\mathcal{Z}}_{0}}{\mathcal{Z}_{-}}}\, ,
\hspace{1cm}
&
k
& = & 
k_{\infty}(f \mathcal{Z}_{+})^{3/4}\, ,
\end{array}
\end{equation}

\noindent 
where the metric function $f$ is given by

\begin{equation}
\label{eq:f}
f^{-3}
=
\tilde{\mathcal{Z}}_{0}\, \mathcal{Z}_{+}\, \mathcal{Z}_{-}\, ,
\end{equation}

\noindent
and the $\mathcal{Z}$ functions take the form

\begin{equation}
\label{eq:Zs}
\begin{array}{rcl}
\tilde{\mathcal{Z}}_{0}
& = &
{\displaystyle
1
+
\frac{\tilde{\mathcal{Q}}_{0}}{\rho^{2}}
+
\frac{2e^{-\phi_{\infty}}k_{\infty}^{2/3}}{3g^{2}}
\frac{\rho^{2} +2\kappa^{2}}{(\rho^{2} +\kappa^{2})^{2}}
\, ,
}
\\
& & \\
\mathcal{Z}_{\pm}
& = &
{\displaystyle1+\frac{\mathcal{Q}_{\pm}}{\rho^{2}}}\, . 
\end{array}
\end{equation}

The mass and entropy of this family of black-hole solutions take
the form

\begin{eqnarray}
\label{eq:mass}
M
& = &
\frac{\pi}{4G_{N}^{(5)}}  
\left[
\tilde{\mathcal{Q}}_{0}
+\frac{2e^{-\phi_{\infty}}k_{\infty}^{2/3}}{3g^{2}}
+\mathcal{Q}_{+} 
+\mathcal{Q}_{-} 
\right]\, ,
\\
& & \nonumber \\
\label{eq:entropy}
S 
& = &
\frac{\pi^{2}}{2G_{N}^{(5)}}  \sqrt{ \tilde{\mathcal{Q}}_{0} \mathcal{Q}_{+}\mathcal{Q}_{-}}\, .  
\end{eqnarray}

Using the charge $\tilde{\mathcal{Q}}_{0}$ instead of $\mathcal{Q}_{0}\equiv
q_{0}/B_{0}$, and assuming that $\tilde{\mathcal{Q}}_{0}$ is not related to
the non-Abelian fields, the mass contains a net $\mathcal{O}(1/g^{2})$
contribution from the instanton while the entropy does not, against the naive
expectations exposed above. We are going to argue that, indeed,
$\tilde{\mathcal{Q}}_{0}$ is a charge completely unrelated to the non-Abelian vector
fields, showing that it counts the number of neutral 5-branes (also known as
solitonic or NSNS 5-branes) while $\mathcal{Q}_{-}$ and $\mathcal{Q}_{+}$
count, respectively, the number of fundamental strings and the momentum along
them. Setting these three charges to zero we are left with the only
non-Abelian component of this solution which is the globally regular and
horizonless gravitating Yang-Mills instanton that we have found in
Ref.~\cite{Cano:2017sqy}, showing that it is is nothing but the dimensional reduction
of Strominger's gauge 5-brane \cite{Strominger:1990et}.

In Ref.~\cite{Cano:2017sqy} we have argued that the gravitating Yang-mills instanton
(or the gauge 5-branes) should not contribute to the entropy while, obviously,
it must contribute to the total mass of black-hole solutions, just as the
global monopole does in 4 dimensions \cite{Hubscher:2008yz,Bueno:2014mea}. The
above mass and entropy formulae reflect this fact.

\section{Embedding in $d=10$ Heterotic Supergravity}
\label{sec-heterotic}

As a first step towards embedding the 5-dimensional supergravity black hole
solution into Heterotic Superstring theory, we are going to embed it in
10-dimensional Heterotic Supergravity ($\mathcal{N}=1,d=10$ supergravity
coupled to vector supermultiplets), with just an SU$(2)$ triplet of gauge
fields. Since the 5-dimensional theory we start from does not have any terms
of higher order in curvatures, we do not consider this kind of terms in the
10-dimensional theory. Observe, however, that the gauge fields occur at first
order in $\alpha'$ and, since our non-Abelian solution has non-trivial vector
fields, in order to be consistent we are forced to study its validity as
solution of Heterotic Superstring theory to first order in $\alpha'$. At this
order there are other terms in the action and we are going to study their
relevance for this solution in Appendix~\ref{app-alpha}.

First of all, we are going to show how the reduction and truncation of the
bosonic sector of the 10-dimensional Heterotic Supergravity with a SU$(2)$
triplet of gauge fields leads to the SU$(2)$-gauged ST$[2,6]$ model of
$\mathcal{N}=1,d=5$ supergravity we are working with.

The action of Heterotic Supergravity in the string frame, including only a
SU$(2)$ triplet of vector fields, is

\begin{equation}
\label{eq:HSaction}
\hat{S}
=
\frac{g_{s}^{2}}{16\pi G_{N}^{(10)}}\int d^{10}x\sqrt{|\hat{g}|}\, 
e^{-2\hat{\phi}}\, 
\left[\hat{R} 
-4(\partial\hat{\phi})^{2}
+\tfrac{1}{2\cdot 3!}\hat{H}^{2}-\alpha' \hat{F}^{A}\hat{F}^{A}\right]\, ,
\end{equation}

\noindent
where the field strengths are defined as

\begin{eqnarray}
\hat{F}^{A}
& = &
d\hat{A}^{A}+\tfrac{1}{2}\epsilon^{ABC}\hat{A}^{B}\wedge\hat{A}^{C}\, ,
\\
& & \nonumber \\
\label{Hdef}
\hat{H}
& = & 
d\hat{B}+2\alpha'\hat{\omega}_{\rm YM}\, ,
\end{eqnarray}

\noindent
and $\omega_{\rm YM}$ is the Chern-Simons 3-form 

\begin{equation}
\omega_{\rm YM}
\equiv 
\hat{F}^{A}\wedge \hat{A}^{A}
-\tfrac{1}{3!}\epsilon^{ABC}\hat{A}^{A}\wedge\hat{A}^{B}\wedge\hat{A}^{C}\, ,
\hspace{1cm}
d\omega_{\rm YM} =  \hat{F}^{A}\wedge \hat{F}^{A}\, . 
\end{equation}

\noindent
In the above expressions, $\alpha'$, the Regge slope, is related to the string
length $\ell_{s}$ by $\alpha'=\ell_{s}^{2}$, and $g_{s}$, the string coupling
constant, is the value of the exponential of the dilaton at infinity:
$g_{s}=e^{\phi_{\infty}}$ in asymptotically-flat configurations. The somewhat
unconventional factor of $g_{s}^{2}$ in front of the action ensures that,
after a rescaling from the string frame to the modified Einstein frame defined
in Ref.~\cite{Maldacena:1996ky} with powers of $e^{\phi-\phi_{\infty}}$, the
action has the standard normalization factor $(16\pi G_{N}^{(10)})^{-1}$. The
10-dimensional Newton constant is given by 

\begin{equation}
G_{N}^{(10)}=8\pi^{6}g_{s}^{2} \ell_{s}^{8}\, .
\end{equation}

If we compactify this theory on $T^{4}$, it is not difficult to see that
truncating all the components of the fields with indices in the internal
coordinates $y^{i}$, $i=1,\cdots,4$, is a consistent truncation. The resulting
6-dimensional action and field strengths have exactly the same form as the
10-dimensional ones, although the action carries an extra factor
$(2\pi\ell_{s})^{4}$ which is the volume of the $T^{4}$:

\begin{equation}
\hat{S}
=
\frac{(2\pi\ell_{s})^{4}g_{s}^{2}}{16\pi G_{N}^{(10)}}\int d^{6}x\sqrt{|\hat{g}|}\, 
e^{-2\hat{\phi}}\, 
\left[\hat{R} 
-4(\partial\hat{\phi})^{2}
+\tfrac{1}{2\cdot 3!}\hat{H}^{2}-\alpha' \hat{F}^{A}\hat{F}^{A}\right]\, .
\end{equation}

The 6-dimensional modified Einstein metric $\hat{g}_{E\, \hat{\mu}\hat{\nu}}$ is
related to the 6-dimensional string metric $\hat{g}_{\hat{\mu}\hat{\nu}}$
by

\begin{equation}
\hat{g}_{\hat{\mu}\hat{\nu}}
= g_{s}^{-1}e^{\hat{\phi}}
\hat{g}_{E\, \hat{\mu}\hat{\nu}}\, ,
\end{equation}

\noindent
and, in this frame, the action takes the form

\begin{equation}
\hat{S}
=
\frac{(2\pi\ell_{s})^{4}}{16\pi G_{N}^{(10)}}\int d^{6}x\sqrt{|\hat{g}_{E}|}\, 
\left[\hat{R}_{E} 
+(\partial\hat{\phi})^{2}
+\tfrac{1}{2\cdot 3!}g_{s}^{2}e^{-2\hat{\phi}}\hat{H}^{2}
-\alpha' g_{s} e^{-\hat{\phi}}\hat{F}^{A}\hat{F}^{A}\right]\, ,
\end{equation}

\noindent
which coincides exactly with the action of the theory of gauged
$\mathcal{N}=(2,0),d=6$ supergravity that we called $\mathcal{N}=2A$ in
Ref.~\cite{Cano:2016rls} upon the redefinitions

\begin{equation}
\hat{\phi}=-\tilde{\varphi}/\sqrt{2}\, , 
\hspace{1cm}
g_{s}\hat{H}/2=\tilde{H}\, ,
\hspace{1cm}
\sqrt{g_{s}\alpha'}\hat{F}^{A}=\tilde{F}^{A}\, ,
\end{equation}

\noindent
which lead to the introduction of the 6-dimensional Yang-Mills coupling
constant $g_{6}= (g_{s}\alpha')^{-1/2}$. 

Further compactification of this theory on a circle leads to the
SU$(2)$-gauged ST$[2,6]$ model of $\mathcal{N}=1,d=5$ supergravity we are
working with, with Newton and Yang-Mills constants given by 

\begin{equation}
G_{N}^{(5)} 
= 
\frac{G_{N}^{(10)}}{(2\pi)^{5}\ell_{s}^{4}R_{z}} 
= 
\frac{\pi g_{s}^{2}\ell_{s}^{4}}{4R_{z}}\, ,
\,\,\,\,\,\,
\mbox{and}
\,\,\,\,\,\,
g = \frac{g_{6}k_{\infty}^{1/3}}{\sqrt{12}} =
\frac{R_{z}^{1/3}}{\sqrt{12 g_{s}\ell^{2}_{s}}}\, .
\end{equation}

\noindent
This reduction was carried out in detail in Ref.~\cite{Cano:2016rls} and we
can use its results, but we have to take into account that we have to rescale
the 5-dimensional metric with the Kaluza-Klein scalar $k$ divided by its
asymptotic value, $k_{\infty}$ in order to preserve the normalization of
asymptotically-flat metrics. This introduces an additional factor of
$k_{\infty}^{1/3}$ in the relations between higher-dimensional fields and
5-dimensional vector fields and an additional factor of $k_{\infty}^{2/3}$ in
the relations between higher-dimensional fields and 5-dimensional 2-form
fields.

Combining the $k_{\infty}$-corrected rules given in Ref.~\cite{Cano:2016rls}
to uplift 5-dimensional configurations to $d=6$ and the relations given above
between 6- and 10-dimensional fields in the string frame, we arrive to the
following rules that allow us to uplift any solution of the SU$(2)$-gauged
ST$[2,6]$ model of $\mathcal{N}=1,d=5$ supergravity to a solution of
10-dimensional Heterotic Supergravity preserving the normalization of the
fields at spatial infinity: 

\begin{equation}
\label{from5to10}
\begin{aligned}
d\hat{s}^{2}
& =
e^{\phi-\phi_{\infty}}
\left[
(k/k_{\infty})^{-2/3}ds^{2} -k^{2}\mathcal{A}^{2}
\right]-dy^{i}dy^{i}\, ,
\\
& \\
\hat{\phi}
& = 
\phi\, ,
\\
& \\
\hat{A}^{A} & =
\frac{k_{\infty}^{1/3}}{\sqrt{12g_{s}\alpha'}}A^{A}
+\frac{\ell^{A}}{\sqrt{\alpha' g_{s}}}\mathcal{A}\, ,
\\
& \\
\hat{H}
& =
-\frac{k_{\infty}^{2/3}}{g_{s} \sqrt{3}}e^{2\phi}k^{-4/3}\star_{(5)}F^{0}
+\frac{k_{\infty}^{1/3}}{g_{s} \sqrt{3}}\mathcal{A}\wedge\mathcal{F}\, ,
\end{aligned}
\end{equation}

\noindent
where we have introduced the auxiliary fields

\begin{equation}
\begin{aligned}
\mathcal{A} 
& \equiv
dz+\frac{k_{\infty}^{1/3}}{\sqrt{12}}A^{+}\, ,
\hspace{1cm}
A^{+} \equiv A^{1}+A^{2}\, ,
\\
& \\
\mathcal{F} 
& \equiv
 F^{-} +\ell^{2}F^{+}+2\ell^{A}F^{A}\, .
\end{aligned}
\end{equation}

Notice that the map gives us the 3-form field strength $\hat{H}$, but not the
2-form potential $\hat{B}$ because the process involves a
dualization. Therefore $\hat{B}$ must be obtained from \req{Hdef} once the
field strengths $\hat{H}$ and $\hat{F}^{A}$ have been computed.

\section{String Theory interpretation}
\label{sec-STinterpretation}

Using the uplifting formulae of the previous section, and defining the
coordinate $u=k_{\infty}z$ (whose period is $2\pi R_{z}$) we get the following
solution of $d=10$ Heterotic Supergravity

\begin{equation}
\label{10dmetric}
\begin{array}{rcl}
d\hat{s}^{2}
& = &
{\displaystyle
\frac{2}{\mathcal{Z}_{-}}du\left(dv-\tfrac{1}{2}\mathcal{Z}_{+}du\right)
-\tilde{\mathcal{Z}}_{0}(d\rho^{2}+\rho^{2}d\Omega_{(3)}^{2})-dy^{i}dy^{i}\, ,
\hspace{1cm}
i=1,2,3,4\, ,
}
\\
& & \\
\hat{B}
& = & 
{\displaystyle
-\frac{1}{\mathcal{Z}_{-}}dv\wedge du
-\tfrac{1}{4}\tilde{\mathcal{Q}}_{0} \cos\theta
d\psi\wedge d\phi\, ,
}
\\
& & \\
\hat{A}^{A}
& = & 
{\displaystyle
-\frac{\rho^{2}}{(\kappa^{2}+\rho^{2})}v^{A}_{L}\, ,
}
\\
& & \\
e^{-2\hat{\phi}}
& = &
{\displaystyle
e^{-2\hat{\phi}_{\infty}}\frac{\mathcal{Z}_{-}}{\tilde{\mathcal{Z}}_{0}}\, ,
}
\end{array}
\end{equation}

\noindent
where $\tilde{\mathcal{Z}}_{0}$ and $\mathcal{Z}_{\pm}$ are given in
Eqs.~(\ref{eq:Zs}). In terms of the stringy constants,
$\tilde{\mathcal{Z}}_{0}$ is given by 

\begin{equation}
\label{eq:tildeZ0}
\tilde{\mathcal{Z}}_{0}
= 
1
+
\frac{\tilde{\mathcal{Q}}_{0}}{\rho^{2}}
+
8\alpha'
\frac{\rho^{2} +2\kappa^{2}}{(\rho^{2} +\kappa^{2})^{2}}
\, .
\end{equation}

\noindent
As shown in Appendix~\ref{app-alpha}, for $\tilde{Q}_{0}>>\kappa^{2}$ this is
a good solution of the Heterotic Superstring effective action to order
$\alpha'$ in the near-horizon ($\rho\rightarrow $) region. This is enough for
our purposes.

Eq.~(\ref{eq:tildeZ0}) shows that the charge $\tilde{\mathcal{Q}}_{0}$ which
is the coefficient of the $1/\rho^{2}$ term is probably associated to
\textit{neutral} (or solitonic or NSNS ) 5-branes \cite{Rey:1989xj} while the
last term should be associated to \textit{gauge} 5-branes. We are first going
to discuss this point in more detail.

We start by noticing that, in absence of the Yang-Mills instanton, this
supergravity solution is the one found in
Refs.~\cite{Tseytlin:1996as,Cvetic:1995bj} which describes solitonic 5-branes
wrapped on $T^{5}$, and fundamental strings wrapped around one cycle of the
$T^{5}$ with momentum along the same direction.

Let us consider the coupling of $N_{S5}$ solitonic 5-branes lying in the
directions $\tfrac{1}{2}(u+v),y^{1},\cdots,y^{4}$, to the Heterotic
Supergravity action given in Eq.~(\ref{eq:HSaction}). Since the effective
action of the solitonic 5-branes is written in terms of the NSNS 6-form
$\tilde{B}$, we must first rewrite the action in terms of that field. It is
convenient to use the language of differential forms, so the action
Eq.~(\ref{eq:HSaction}) takes the form

\begin{equation}
\hat{S}
=
\frac{g_{s}^{2}}{16\pi G_{N}^{(10)}}\int
e^{-2\hat{\phi}}\, 
\left[\star \hat{R} 
-4d\hat{\phi}\wedge \star d\hat{\phi}
+\tfrac{1}{2}\hat{H}\wedge \star \hat{H}
+2\alpha' \hat{F}^{A}\wedge \star\hat{F}^{A}\right]\, ,
\end{equation}

\noindent
and, after dualization $\star e^{-2\hat{\phi}}\hat{H}= \hat{\tilde{H}}\equiv
d\hat{\tilde{B}}$

\begin{equation}
\begin{array}{rcl}
\hat{S}
& = &
{\displaystyle
\frac{g_{s}^{2}}{16\pi G_{N}^{(10)}}\int
}
\left\{
e^{-2\hat{\phi}}\,
\left[\star \hat{R} 
-4d\hat{\phi}\wedge \star d\hat{\phi}
+2\alpha' \hat{F}^{A}\wedge \star\hat{F}^{A}\right]
\right.
\\
& & \\
& & 
\left.
+\tfrac{1}{2}e^{2\hat{\phi}}\hat{\tilde{H}}\wedge \star \hat{\tilde{H}}
+2\alpha' \hat{\tilde{B}}\wedge \hat{F}^{A}\wedge \hat{F}^{A}
\right\}\, .
\end{array}
\end{equation}

The 6-form will couple to the Wess-Zumino term in the effective action of
$N_{S5}$ coincident solitonic 5-branes via its pullback over the worldvolume

\begin{equation}
N_{S5}T_{S5}\, g_{s}^{2}\int \phi_{*}\hat{\tilde{B}} \, , 
\hspace{.8cm}
\mbox{where}
\hspace{.8cm}
T_{S5}= \frac{1}{(2\pi\ell_{s})^{5}\ell_{s} g_{s}^{2}}\, ,
\end{equation}

\noindent
and the 6-form equation of motion is

\begin{equation}
\frac{g_{s}^{2}}{16\pi G_{N}^{(10)}}
\left\{
d(\star e^{2\hat{\phi}}\hat{\tilde{H}}) -2\alpha' \hat{F}^{A}\wedge
\hat{F}^{A}
\right\}
=
g_{s}^{2}N_{S5}T_{S5}\star_{(4)}\delta^{(4)}(\rho)\, ,
\end{equation}

\noindent
where $\star_{(4)}\delta^{(4)}(\rho)$ is a 4-form in the 5-branes' transverse
space whose integral gives $1$. 

Integrating both sides of this equation over the transverse space\footnote{We
  replace $\star e^{2\hat{\phi}}\hat{\tilde{H}}$ by $\hat{H}$ for simplicity
  and use Stokes' theorem in the first term. For the second term we have
\begin{equation}
\frac{1}{16\pi^{2}}\int_{\mathbb{R}^{4}}\hat{F}^{A}\wedge \hat{F}^{A}=1\, ,
\end{equation}
the instanton number.
}  we get 

\begin{equation}
\tilde{\mathcal{Q}}_{0}=\mathcal{Q}_{0} -8\alpha^{\prime} = \ell_{s}^{2}N_{S5}\, ,
\end{equation}

\noindent
which confirms that $\tilde{\mathcal{Q}}_{0}=\mathcal{Q}_{0}-8\alpha' n$,
where $n$ would the instanton number in more general configurations counts
solitonic (neutral) 5-branes. The number of gauge 5-branes $N_{G5}$ coincides
with the instanton number $n$. Thus, we conclude that the parameter
$\mathcal{Q}_{0}$ of the solution is

\begin{equation}
\mathcal{Q}_{0}= \ell_{s}^{2}(N_{S5}+8N_{G5})\, .
\end{equation}

The function $\mathcal{Z}_{-}$ is clearly associated to 10-dimensional
fundamental strings wrapped around the coordinate $\tfrac{1}{2}(u-v)$.  If we
couple $N_{F1}$ fundamental strings lying in the direction $\tfrac{1}{2}(u-v)$
we have

\begin{equation}
T_{F1}N_{F1} 
= 
\frac{g_{s}^{2}}{16\pi G_{N}^{(10)}}\int_{V^{8}} d(\star e^{-2\hat{\phi}}
\hat{H})\, , 
\hspace{.8cm}
\mbox{where}
\hspace{.8cm}
T_{F1} = \frac{1}{2\pi\alpha'}\, ,
\end{equation}

\noindent
where $V^{8}$ is the space transverse to worldsheet parametrized by $u$ and
$v$, whose boundary is the product $T_{4}\times S^{3}_{\infty}$. Using Stokes'
theorem and the value of volume of $T^{4}$ $(2\pi \ell_{S})^{4}$, we get

\begin{equation}
\mathcal{Q}_{-} = \ell_{s}^{2}g_{s}^{2}N_{F1}\, .  
\end{equation}

Finally, the function $\mathcal{Z}_{+}$ is associated to a gravitational wave
moving in the compact direction $\tfrac{1}{2}(u-v)$ at the speed of light. The
simplest way to compute its momentum is to T-dualize the solution along that
direction. This operation interchanges winding number ($N_{F1}$) and momentum
($N_{W}$) and, at the level of the solution, it interchanges the functions
$\mathcal{Z}_{-}$ and $\mathcal{Z}_{+}$ or, equivalently, the constants
$\mathcal{Q}_{-}$ and $\mathcal{Q}_{+}$. Thus,

\begin{equation}
\mathcal{Q}_{+} 
= 
\ell_{s}^{2}g_{s}^{\prime\, 2}N'_{F1} 
= 
\ell_{s}^{2} \left(g_{s}\ell_{s}/R_{z} \right)^{2}N_{W}
=
\frac{g_{s}^{2}\ell^{4}}{R_{z}^{2}}N_{W}\, ,
\end{equation}

\noindent
where we have taken into account the transformation of the string coupling
constant under T-duality.

We conclude that the fields that give rise to the 5-dimensional non-Abelian
black hole in Eq.~(\ref{eq:3chargebh1}),(\ref{eq:f}) and (\ref{eq:Zs})
correspond to those sourced by $N_{F1}$ fundamental strings wrapped around the
6th dimension with $N_{W}$ units of momentum moving in the same direction and
$N_{S5}$ solitonic (neutral) and $N_{G5}=1$ gauge 5-branes wrapped around the
6th direction and a $T^{4}$. In terms of these numbers, the black hole's mass
and the entropy in Eqs.~(\ref{eq:mass}) and (\ref{eq:entropy}) take the form

\begin{eqnarray}
M 
& = &
\frac{R_{z}}{g^{2}_{s}\ell_{s}^{2}}(N_{S5}+8N_{G5})
+
\frac{R_{z}}{\ell_{s}^{2}}N_{F1}
+
\frac{1}{R_{z}}N_{W}\, ,
\\
& & \nonumber \\
S & = & 2\pi \sqrt{N_{F1}N_{W}N_{S5}}\, .
\end{eqnarray}

Unfortunately, the dynamics of String Theory in the background of
non-perturbative objects such as solitonic and gauge 5-branes is not as well
understood as its dynamics in the background of D-branes. Therefore, it is
convenient to perform a string-weak coupling Heterotic-Type-I duality
transformation \cite{Dabholkar:1995ep,Hull:1995nu,Polchinski:1995df} which
acts on the fields as follows:\footnote{These are the transformations that
  preserve the normalization of the string metric at spatial infinity and lead
  to the correct normalization of the action of the Type-I theory. In
  particular, the rescaling of the gauge fields is required in order to
  reproduce correctly the term that appears in the expansion of the
  Born-Infeld action of the O9-D9-brane system (in the Abelian case). The
  effective worldvolume action of the D9-brane (Born-Infeld plus Wess-Zumino
  (WZ) terms) is
\begin{equation}
\hat{S}_{D9}
=
T_{D9}g_{I}\int d\xi^{10} e^{-\hat{\varphi}}
\sqrt{\mathrm{det}(\hat{\jmath}_{ij}
+2\pi\alpha' \hat{\mathcal{F}}_{ij})} +WZ\, ,
\end{equation}
where $g_{I}$ is the Type~I string coupling constant.  In the physical gauge,
ignoring the cosmological constant-type term because it will be cancelled by
the O9-planes, and using $T_{D9}=[(2\pi\ell_{s})^{9}\ell_{s}g_{I}]^{-1}$ we
get
\begin{equation}
\hat{S}_{D9}
\sim
\frac{g_{I}^{2}}{16\pi G_{N,I}^{(10)}}
\int d^{10}x \sqrt{|\hat{\jmath}|}\,\left[\alpha^{\prime} e^{-\hat{\varphi}}
  \hat{\mathcal{F}}^{2}\right] +WZ\, ,  
\end{equation}
where, now, $16\pi G_{N,I}^{(10)} = (2\pi \ell_{s})^{7}\ell_{s}g_{I}^{2}$.  If
we rewrite the Type-I supergravity action in terms of the RR 6-form
$\hat{C}^{(6)}$, just as in the Heterotic case, we get a term
$\hat{C}^{(6)}\wedge \hat{\mathcal{F}}^{A}\wedge \hat{\mathcal{F}}^{A}$. This
term originates in the WZ term of the D9 effective action as well.
}${}^{,}$\footnote{The same procedure (a strong-weak coupling duality
  transformation within Type-IIB supergravity) was followed in
  Ref.~\cite{Callan:1996dv} to derive the D5D1W solution without non-Abelian
  fields from the solution in
  \cite{Tseytlin:1996as,Tseytlin:1996as,Cvetic:1995bj} which can be embedded 
  directly in the Type-IIB NSNS sector. The presence of non-Abelian vector
  fields suggests the route we have taken.}

\begin{equation}
\hat{g}_{\hat{\mu}\hat{\nu}} 
= 
e^{-(\hat{\varphi}-\hat{\varphi}_{\infty})}
\hat{\jmath}_{\hat{\mu}\hat{\nu}}\, ,
\hspace{.6cm}
\hat{\phi} 
= 
-\hat{\varphi}\, ,
\hspace{.6cm}
\hat{C}^{(2)}_{\hat{\mu}\hat{\nu}}
=
e^{-\hat{\varphi}_{\infty}}\hat{B}_{\hat{\mu}\hat{\nu}}  
\hspace{.6cm}
\hat{A}^{A}_{\hat{\mu}}
=
g_{I}^{1/2}\hat{\mathcal{A}}^{A}_{\hat{\mu}}\, ,
\end{equation}

\noindent
where $g_{I}\equiv e^{\hat{\varphi}_{\infty}}$ is the Type-I string coupling
constant. These transformations lead to the Type-I supergravity action

\begin{equation}
\begin{array}{rcl}
g_{I}^{-4}\hat{S}_{I}
& = &
{\displaystyle
\frac{g_{I}^{2}}{16\pi G_{N,I}^{(10)}}\int
}
\left\{
e^{-2\hat{\varphi}}\,
\left[\star \hat{R} 
-4d\hat{\varphi}\wedge \star d\hat{\varphi}
\right]
+\tfrac{1}{2}\hat{G}^{(3)}\wedge \star \hat{G}^{(3)}
+2\alpha' e^{-\hat{\varphi}}\hat{\mathcal{F}}^{A}\wedge \star\hat{\mathcal{F}}^{A}
\right\}\, ,
\end{array}
\end{equation}

\noindent
and our solution takes the form

\begin{equation}
\label{eq:typeIsolution}
\begin{array}{rcl}
d\hat{s}_{I}^{2}
& = &
{\displaystyle
\frac{2}{\sqrt{\tilde{\mathcal{Z}}_{0}\mathcal{Z}_{-}}}
du\left(dv-\tfrac{1}{2}\mathcal{Z}_{+}du\right)
-\sqrt{\tilde{\mathcal{Z}}_{0}\mathcal{Z}_{-}}
(d\rho^{2}+\rho^{2}d\Omega_{(3)}^{2})
-\sqrt{\frac{\mathcal{Z}_{-}}{\tilde{\mathcal{Z}}_{0}}}dy^{i}dy^{i}\, ,
}
\\
& & \\
\hat{C}^{(2)}
& = & 
{\displaystyle
-\frac{e^{-\hat{\varphi}_{\infty}}}{\mathcal{Z}_{-}}dv\wedge du
-\frac{e^{-\hat{\varphi}_{\infty}}}{4}\tilde{\mathcal{Q}}_{0}  \cos\theta
d\psi\wedge d\phi\, ,
}
\\
& & \\
\hat{\mathcal{A}}^{A}
& = & 
{\displaystyle
-e^{-\hat{\varphi}_{\infty}/2}\frac{\rho^{2}}{(\kappa^{2}+\rho^{2})}v^{A}_{L}\, ,
}
\\
& & \\
e^{-2\hat{\varphi}}
& = &
{\displaystyle
e^{-2\hat{\varphi}_{\infty}}\frac{\tilde{\mathcal{Z}}_{0}}{\mathcal{Z}_{-}}\, .
}
\end{array}
\end{equation}

In agreement with the fact that under Heterotic/Type-I duality fundamental
strings and solitonic 5-branes transform into D1- and D5-branes, respectively,
gravitational waves remain gravitational waves with the same momentum, this
solution describes the fields produced by a D5-brane intersecting a D1-brane
in the $z$ direction with a wave propagating along that direction. The
Yang-Mills instanton is a non-perturbative configuration of the non-Abelian
Born-Infeld field that occurs in the worldvolume of the parallel D9-branes
that give rise to the Type-I theory from the Type-IIB and sources D5-branes.
Thus $N_{D1}=N_{F1}$, $N_{D5}=N_{S5}$, $N_{GD5}=N_{G5}$ and, in Type-I
variables, the mass and entropy formulae take the form

\begin{eqnarray}
M 
& = &
\frac{R_{z}}{g_{I}\ell_{s}^{2}}(N_{D5}+8N_{GD5})
+
\frac{R_{z}}{g_{I}\ell_{s}^{2}}N_{D1}
+
\frac{1}{R_{z}}N_{W}\, ,
\\
& & \nonumber \\
S & = & 2\pi \sqrt{N_{D1}N_{D5}N_{W}}\, .
\end{eqnarray}

In absence of the instanton ($N_{GD5}=0$) this solution is identical to the
one originally considered in Ref.~\cite{Callan:1996dv}, which is itself very
closely related to Strominger and Vafa's original model
\cite{Strominger:1996sh}.\footnote{See also
  Refs.~\cite{Maldacena:1996ky,Peet:2000hn,David:2002wn}.}  The same
conditions (namely, that all the $N$s are large and $N_{W}>> N_{D1,D5}$)
ensure that this solution describes at leading order in $\alpha'$ (low
curvature) and in $g_{s}$ (perturbative string theory) a good background for
Type-IIB string theory.

\section{Discussion}
\label{sec-discussion}

In the previous sections we have shown that the 5-dimensional supergravity
black holes with 3 quantized Abelian charges $N_{D1},N_{D5},N_{W}$ and a
non-Abelian instanton can be seen, up to dualities, as the fields associated
to a 10-dimensional Type-IIB configuration with

\begin{enumerate}
\item An orientifold O9$_{+}$-plane and 16 D9-branes and their mirror images,
  that give rise to the Type-I superstring theory with gauge group SO$(32)$
  (see, \textit{e.g.}~\cite{Angelantonj:2002ct} and references therein).
\item $N_{D5}$ D5-branes wrapped around the 5th-9th directions and $N_{D1}$
  D-strings wrapped around the 5th direction with $N_{W}$ units of momentum
  along the 5th direction. Open strings can end on these D-strings and
  D5-branes. 
\item $N_{GD5}=1$ ``gauge D5-brane'', sourced by an instanton field located in
  the 1st-4th dimensions, which are not compact. This brane, which is the dual
  of the heterotic gauge 5-brane is often referred to as a D5-brane
  ``dissolved'' into the spacetime-filling D9-branes and differs essentially
  from standard D5-branes because no strings can end on them.
 \end{enumerate}

 Since the entropy of the D1D5W black holes can be understood as associated to
 the massless states associated to strings with one endpoint on a D1 and the
 other on a D5 (1-5 states) and this fact, as discussed  in
 Ref.~\cite{Callan:1996dv} is unchanged by the presence of the D9-branes and
 O9$_{+}$-plane that defines the Type-I theory\footnote{The counting of states
   is, however, different since, as mentioned in Ref.~\cite{Callan:1996dv} one
   has to take into account the SU$(2)$ degrees of freedom associated to the
   D5-brane of the Type-I string found in \cite{Witten:1995gx}.} the
 microscopic interpretation of the entropy of these non-Abelian black holes
 must be the same as in the Abelian case and should give the same result at
 leading order.  Observe that, as an intermediate step in the uplift of the
 solution to 10 dimensions one obtains a non-Abelian string solution in 6
 dimensions with an AdS$_{3}\times$S$_{3}$ near-horizon geometry where the
 AdS$_{3}$ radius only depends on 3 quantized Abelian charges
 $N_{D1},N_{D5},N_{W}$.

 It is important to stress that the correct identification of the charges and
 their meaning in terms of branes plays a crucial r\^ole to reach this
 conclusion as well as in solving the apparent non-Abelian hair problem
 explained in the Introduction. A more detailed study is, however, necessary
 to find corrections to the entropy. In particular, the $\alpha'$ corrections
 to this solution in the asymptotic limit need to be determined (see the
 Appendix).

 In the last few years we have constructed non-Abelian static and rotating
 black-hole solutions in 4 and 5 dimensions
 \cite{Huebscher:2007hj,Meessen:2008kb,Hubscher:2008yz,Bueno:2014mea,Meessen:2015nla,Meessen:2015enl},
 as well as black-ring solutions \cite{Ortin:2016bnl} and microstate geometries \cite{Ramirez:2016tqc} in 5 dimensions. All
 those constructed with ``colored monopoles'' in 4 dimensions and many of the
 5-dimensional solutions exhibit non-Abelian hair which seems to contribute to
 the entropy or the angular momentum on the horizon but cannot be seen at
 infinity.  Many of them can be uplifted to 10-dimensional Heterotic
 Supergravity and then dualized into Type-I Supergravity solutions and it is
 likely that the correct interpretation of the charges of those solutions is
 enough to understand the non-Abelian hair problem. Work in this direction is
 in progress.

\section*{Acknowledgments}

TO would like to thank \'Angel Uranga for very useful conversations.  This
work has been supported in part by the Spanish Government grants
FPA2012-35043-C02-01 and FPA2015-66793-P (MINECO/FEDER, UE), the Centro de
Excelencia Severo Ochoa Program grant SEV-2012-0249 and the Spanish
Consolider-Ingenio 2010 program CPAN CSD2007-00042.  The work of PAC was
supported by a ``la Caixa-Severo Ochoa'' International pre-doctoral grant. The
work of PFR was supported by the \textit{Severo Ochoa} pre-doctoral grant
SVP-2013-067903.  TO wishes to thank M.M.~Fern\'andez for her permanent
support.

\appendix

\section{The issue of $\alpha'$ corrections}
\label{app-alpha}

As we have mentioned in the main body of the paper, the solution of
10-dimensional Heterotic Supergravity that we have obtained by uplifting the
5-dimensional non-Abelian supersymmetric black hole solution has non-trivial
SU$(2)$ gauge fields. These occur at first order in $\alpha'$ in the
low-energy Heterotic Superstring effective action together with other terms
that we are going to describe following Ref.~\cite{Bergshoeff:1989de}, and
which we have ignored. The purpose of this appendix is to study the relevance
of the omitted terms for the solution at hands. Only if these are negligible
with respect to those we have considered can the solution be considered a good
solution of the Heterotic Superstring effective action to first order in
$\alpha'$.

At lowest order (zeroth) in $\alpha'$, the Heterotic Superstring effective
action is nothing but the action of pure $\mathcal{N}=1,d=10$ supergravity
\cite{Chamseddine:1980cp,Bergshoeff:1981um}.  The coupling to super-Yang-Mills
multiplets \cite{Bergshoeff:1981um,Chapline:1982ww} leads to the exactly
supersymmetric Heterotic Supergravity theory described in
Section~\ref{sec-heterotic}. From the point of view of the Heterotic
Superstring effective action, the terms associated to the Yang-Mills fields
are of higher order in $\alpha'$: their kinetic term occurs in the action
Eq.~(\ref{eq:HSaction}) at first order and their Chern-Simons 3-form
$\omega_{\rm YM}$ occurs in the Kalb-Ramond 3-form field strength $\hat{H}$ at
first order as well, Eq.~(\ref{Hdef}), modifying its Bianchi identity so that
it takes the form

\begin{equation}
d\hat{H}  = 2\alpha'  \hat{F}^{A}\wedge \hat{F}^{A}\, .
\end{equation}

This correction in $\hat{H}$ introduces terms of second order in $\alpha'$ in
the action and in the Einstein equations but it is precisely this mixture of
terms of different orders in $\alpha'$ that is exactly supersymmetric and
gauge invariant.

The existence of additional terms at first order in $\alpha'$ in the Heterotic
Superstring effective action is both a blessing, because it makes possible the
Green-Schwarz anomaly-cancellation mechanism \cite{Green:1984sg}, and a curse
because, once they are included, the action will only be supersymmetric and
gauge-invariant to second order in $\alpha'$ \cite{Bergshoeff:1988nn}. The
addition of further $\alpha'$ corrections only makes the action supersymmetric
and gauge-invariant to higher order in $\alpha'$ \cite{Bergshoeff:1989de}
and will not be considered here.

With the addition of the aforementioned missing terms, the Heterotic
Superstring effective action takes the form

\begin{equation}
\label{eq:HSaction+corrections}
\hat{S}
=
\frac{g_{s}^{2}}{16\pi G_{N}^{(10)}}\int d^{10}x\sqrt{|\hat{g}|}\, 
e^{-2\hat{\phi}}\, 
\left\{
\hat{R} 
-4(\partial\hat{\phi})^{2}
+\tfrac{1}{2\cdot 3!}\hat{H}^{2}
-\alpha' 
\left[
\hat{F}^{A}\hat{F}^{A} 
+\hat{R}_{(-)}{}^{\hat{a}}{}_{\hat{b}}\hat{R}_{(-)}{}^{\hat{b}}{}_{\hat{a}} 
\right]
\right\}\, ,
\end{equation}

\noindent
where 

\begin{equation}
\hat{R}_{(-)}{}^{\hat{a}}{}_{\hat{b}}\hat{R}_{(-)}{}^{\hat{b}}{}_{\hat{a}}
=
\hat{R}_{(-)\, \hat{\mu}\hat{\nu}}{}^{\hat{a}}{}_{\hat{b}}
\hat{R}_{(-)}{}^{\hat{\mu}\hat{\nu}\,  \hat{b}}{}_{\hat{a}} \, .
\end{equation}

Here $\hat{\Omega}_{(-)}{}^{\hat{a}}{}_{\hat{b}}$ is one of the two
\textit{torsionful spin connection} 1-forms that can be constructed by adding
to the Levi-Civita spin connection $\hat{\omega}^{\hat{a}\hat{b}}$ 1-form a
torsion piece

\begin{equation}
\hat{\Omega}_{(\pm)}{}^{\hat{a}}{}_{\hat{b}} 
=
\hat{\omega}^{\hat{a}}{}_{\hat{b}}
\pm
\tfrac{1}{2}\hat{H}_{\hat{\mu}}{}^{\hat{a}}{}_{\hat{b}}dx^{\hat{\mu}}\, ,
\end{equation}

\noindent
whose curvature 2-forms are defined by

\begin{equation}
\hat{R}_{(\pm)}{}^{\hat{a}}{}_{\hat{b}}
= 
d \hat{\Omega}_{(\pm)}{}^{\hat{a}}{}_{\hat{b}}
- \hat{\Omega}_{(\pm)}{}^{\hat{a}}{}_{\hat{c}}\wedge  
\hat{\Omega}_{(\pm)}{}^{\hat{c}}{}_{\hat{b}}\, .
\end{equation}

\noindent
The Kalb-Ramond field strength 3-form is now given by 

\begin{equation}
\hat{H}
= 
d\hat{B}
+2\alpha'\left(\hat{\omega}_{\rm YM}
+\hat{\omega}_{{\rm L}\, (-)}\right)\, , 
\end{equation}

\noindent
where $\hat{\omega}_{{\rm L}\, (\pm)}$ is the Chern-Simons 3-form of the
torsionful spin connection $\hat{\Omega}_{(\pm)}$

\begin{equation}
\hat{\omega}_{{\rm L}\, (\pm)}
= 
d\hat{\Omega}_{(\pm)}{}^{\hat{a}}{}_{\hat{b}} \wedge 
\hat{\Omega}_{(\pm)}{}^{\hat{b}}{}_{\hat{a}} 
-\tfrac{2}{3}
\hat{\Omega}_{(\pm)}{}^{\hat{a}}{}_{\hat{b}} \wedge 
\hat{\Omega}_{(\pm)}{}^{\hat{b}}{}_{\hat{c}} \wedge
\hat{\Omega}_{(\pm)}{}^{\hat{c}}{}_{\hat{a}}\, ,
\end{equation}

\noindent
and, correspondingly, its Bianchi identity becomes

\begin{equation}
d\hat{H}  
= 
2\alpha'  \left(\hat{F}^{A}\wedge \hat{F}^{A} 
+\hat{R}_{(-)}{}^{\hat{a}}{}_{\hat{b}}\wedge \hat{R}_{(-)}{}^{\hat{b}}{}_{\hat{a}}\right)\, .
\end{equation}

Written in this way, and besides the explicit ones, the action contains an
infinite number of implicit $\alpha'$ corrections which arise due to the
recursive way in which $\hat{H}$ is defined, since it depends on the
Chern-Simons form of $\hat{\Omega}_{(-)}$, which is defined in terms of
$\hat{H}$. At the order at which we are working, it is enough to
keep in the definitions of $\hat{\Omega}_{(\pm)}$ only the terms of zeroth
order in $\alpha'$, that is

\begin{equation}
\hat{\Omega}_{(\pm)}{}^{\hat{a}}{}_{\hat{b}} 
=
\hat{\omega}^{\hat{a}}{}_{\hat{b}}
\pm
\tfrac{1}{2}\hat{H}^{(0)}_{\hat{\mu}}{}^{\hat{a}}{}_{\hat{b}}dx^{\hat{\mu}}\, ,
\,\,\,\,\,
\mbox{where}
\,\,\,\,\,
\hat{H}^{(0)} \equiv d\hat{B}\, ,
\end{equation}

\noindent
and we will ignore all the $\alpha'^2$ terms in the action
Eq.~(\ref{eq:HSaction+corrections}).

Now, by plugging the solution Eq.~(\ref{10dmetric}) into the equations of
motion that follow from the action Eq.~(\ref{eq:HSaction+corrections}) with
the torsionful spin connection defined in the previous equation, we can study
if they are satisfied to first order in $\alpha'$. 

Following the scheme explained in Ref.~\cite{Bergshoeff:1992cw}
the variations of the action with respect to the each field can be separated
into variations with respect to explicit occurrence of the field in the action
and variations with respect to the implicit occurrence via the torsionful spin
connection. The former are the zeroth order equations plus terms proportional
to the so-called ``$\hat{T}$-tensors,'' which we will define shortly and are
of order $\alpha'$. According to the lemma proved in Section~3 of
Ref.~\cite{Bergshoeff:1989de}, the latter are of order $\alpha'$ and
proportional to the zeroth order equations of motion. Since the solution
Eq.~(\ref{10dmetric}) satisfies the zeroth order equations of motion up to
terms of first order in $\alpha'$, the implicit variations are of order
$\alpha'^2$ and can be ignored.

The conclusion is that it is enough to study the $\hat{T}$-tensor-corrected
zeroth-order equations of motion. The 3 $\hat{T}$-tensors that appear in the
corrections are defined as

\begin{equation}
\begin{array}{rcl}
\hat{T}_{\hat{\mu}\hat{\nu}\hat{\rho}\hat{\sigma}}
& \equiv &
\alpha'\left[
\hat{F}_{[\hat{\mu}\hat{\nu}}{}^{A}
\hat{F}_{\hat{\rho}\hat{\sigma}]}{}^{A}
+
\hat{R}_{(-)\, [\hat{\mu}\hat{\nu}|}{}^{\hat{a}}{}_{\hat{b}}
\hat{R}_{(-)\, |\hat{\rho}\hat{\sigma}]}{}^{\hat{b}}{}_{\hat{a}}
\right]\, ,
\\
& & \\ 
\hat{T}_{\hat{\mu}\hat{\nu}}
& \equiv &
\alpha'\left[
\hat{F}_{\hat{\mu}\hat{\rho}}{}^{A}\hat{F}_{\hat{\nu}}{}^{\hat{\rho}\, A} 
+
\hat{R}_{(-)\, \hat{\mu}\hat{\rho}}{}^{\hat{a}}{}_{\hat{b}}
\hat{R}_{(-)\, \hat{\nu}}{}^{\hat{\rho}\,  \hat{b}}{}_{\hat{a}}
\right]\, ,
\\
& & \\    
\hat{T}
& \equiv &
\hat{T}^{\hat{\mu}}{}_{\hat{\mu}}\, .
\\
\end{array}
\end{equation}

The 4-form $\hat{T}$-tensor is the r.h.s.~of the Bianchi identity of
$\hat{H}$, the symmetric 2-index $\hat{T}$-tensor is the term that occurs in
the Einstein equations and its trace occurs in the dilaton equation.

The Yang-Mills part of these tensors was included in the equations of motion
of the Heterotic Supergravity Eq.~(\ref{eq:HSaction}). Therefore, we just need
to compute them and compare the Lorentz curvature part with the Yang-Mills
part. In other words, we need to compare the $\kappa$-dependent term with the
rest, which should be much smaller.\footnote{It is worth stressing this point:
  since our starting point is not an exact solution of the action to zeroth
  order in $\alpha'$, our goal is not to make the value of the
  $\hat{T}$-tensors as small as possible.}

For the solution at hands, to $\mathcal{O}(\alpha'^{2})$, they are explicitly
given by

\begin{eqnarray}
\hat{T}^{(4)}
& \sim & 
\alpha'\left[\frac{\kappa^{4}}{(\kappa^{2}+\rho^{2})^{4}}
-\frac{\tilde{Q}_{0}^{2}}{(\tilde{Q}_{0}+\rho^{2})^{4}}\right]
d\rho \rho^{3}\wedge \sin{\theta} d\theta\wedge d\Psi\wedge d\phi
\, ,
\\
& & \nonumber \\
\hat{T}_{uu}
& = &
- \alpha'\frac{32 Q_{-} Q_{+} \rho^{4} \left[\tilde{Q}_{0}^{2}
+\tilde{Q}_{0} \left(Q_{-}+3 \rho^{2}\right)+Q_{-}^{2}+3 Q_{-} \rho^{2}
+3\rho^{4}\right]}{\left(\tilde{Q}_{0}+\rho^{2}\right)^{4} 
\left(Q_{-}+\rho^{2}\right)^{4}}\, ,
\\
& & \nonumber \\
\hat{T}_{ij}
& = &
 \alpha'\delta_{ij}\frac{48 \rho^{2} }{\left(\tilde{Q}_{0}+\rho^{2}\right)^5} 
\left[\tilde{Q}_{0}^{2}-\frac{\kappa^{4} \left(\tilde
Q_{0}+\rho^{2}\right)^{4}}{\left(\kappa^{2}+\rho^{2}\right)^{4}}\right]\, ,
\\
& & \nonumber \\
\hat{T}
& = &
- \alpha' \frac{192 \rho^{4} 
}{\left(\kappa^{2}+\rho^{2}\right)^{4}
\left(\tilde{Q}_{0}+\rho^{2}\right)^{6}}
\left[\kappa^{8} \tilde{Q}_{0}^{2}+4 \kappa^{6} \tilde{Q}_{0}^{2} \rho^{2}
\right.
\nonumber \\
& & \nonumber \\
& & 
\left.
-\kappa^{4} \left(\tilde{Q}_{0}^{4}+4 \tilde{Q}_{0}^{3}
\rho^{2}+4 \tilde{Q}_{0} \rho^{6}+\rho^{8}\right)
+4 \kappa^{2} \tilde{Q}_{0}^{2}
\rho^{6} +\tilde{Q}_{0}^{2} \rho^8\right]
\end{eqnarray}

\noindent
where $\hat{T}^{(4)} =
\frac{1}{4!}\hat{T}_{\hat{\mu}\hat{\nu}\hat{\rho}\hat{\sigma}}
dx^{\hat{\mu}}dx^{\hat{\nu}}dx^{\hat{\rho}}dx^{\hat{\sigma}}$ and
$i,j=2,3,4,5$ label the 4 coordinates of the 5-branes worldvolume which are
orthogonal to the wave.

Let us start by analyzing $\hat{T}^{(4)}$: in the near-horizon region
$\rho\rightarrow 0$ the leading term is 

\begin{equation}
\hat{T}^{(4)}
\sim 
\alpha'\left(\frac{1}{\kappa^{4}}-\frac{1}{\tilde{Q}_{0}^{2}} \right)
d\rho \rho^{3}\wedge \sin{\theta} d\theta\wedge d\Psi\wedge d\phi
\, .  
\end{equation}

\noindent
In this limit, the $\alpha'$ corrections of our solution will be small if
$\kappa^{-4}>> \tilde{Q}_{0}^{-2}$, that is, if $\tilde{Q}_{0}>>\kappa^{2}$ so
the number of S5-branes is very large.

Asymptotically ($\rho\rightarrow \infty$), the leading term is 

\begin{equation}
\hat{T}^{(4)}
\sim 
\alpha'\frac{(\kappa^{4}-\tilde{Q}_{0}^{2})}{\rho^{8}}
d\rho \rho^{3}\wedge \sin{\theta} d\theta\wedge d\Psi\wedge d\phi
\, ,  
\end{equation}

\noindent
and the absence of $\alpha'$ corrections in this limit requires exactly the
opposite to happen: $\tilde{Q}_{0}<<\kappa^{2}$ and the number of S5-branes
should be very small.

The analysis of the other tensors sheds identical results, indicating that we
can only consider our solution a good solution of the Heterotic Superstring
effective action in either the near-horizon $\rho\rightarrow 0$ region for
$\tilde{Q}_{0}>>\kappa^{2}$ or in the asymptotic $\rho\rightarrow 0$ region
for $\tilde{Q}_{0}<<\kappa^{2}$. In either case, the solution will have to be
$\alpha'$ corrected in the other region.

For the purpose of computing the entropy it is more convenient to take
$\tilde{Q}_{0}>>\kappa^{2}$ so that the near-horizon region is well described
to order $\alpha'$ in Heterotic Superstring effective action. The $\alpha'$
corrections which are needed in the asymptotic limit will be determined and
studied in a forthcoming publication \cite{kn:CMOR}.



\begin{thebibliography}{99}

\bibitem{Smoller:1991ag}
J.~A.~Smoller, A.~G.~Wasserman, S.~T.~Yau and J.~B.~McLeod,
``Smooth static solutions of the Einstein Yang-Mills equations,''
Commun.\ Math.\ Phys.\  {\bf 143} (1991) 115.
\doi{10.1007/BF02100288}

\bibitem{Galtsov:1989ip}
D.~V.~Galtsov and A.~A.~Ershov,
``Nonabelian Baldness of Colored Black Holes,''
Phys.\ Lett.\ A {\bf 138} (1989) 160.
\doi{10.1016/0375-9601(89)90019-4}

\bibitem{Ershov:1991nv}
A.~A.~Ershov and D.~V.~Galtsov,
``Nonexistence of regular monopoles and dyons in the SU(2) Einstein Yang-Mills theory,''
Phys.\ Lett.\ A {\bf 150} (1990) 159.
\doi{10.1016/0375-9601(90)90113-3}

\bibitem{Bizon:1992pi}
P.~Bizon and O.~T.~Popp,
``No hair theorem for spherical monopoles and dyons in SU(2) Einstein Yang-Mills theory,''
Class.\ Quant.\ Grav.\  {\bf 9} (1992) 193.
\doi{10.1088/0264-9381/9/1/017}

\bibitem{Volkov:1998cc}
M.~S.~Volkov and D.~V.~Gal'tsov,
``Gravitating nonAbelian solitons and black holes with Yang-Mills fields,''
Phys.\ Rept.\  {\bf 319} (1999) 1.
\doi{10.1016/S0370-1573(99)00010-1}
[\hepth{9810070}].

\bibitem{Galtsov:2001myk}
D.~V.~Gal'tsov,
``Gravitating lumps,''
Proceedings of the 16th International Conference on General Relativity 
and Gravitation (GR16), 
Nigel T. Bishop and Sunil D. Maharaj Eds.,
World Scientific,  Singapore, 2002.
\hepth{0112038}.

\bibitem{Volkov:1989fi}
M.S.~Volkov and D.V.~Galtsov,
``NonAbelian Einstein Yang-Mills black holes,''
JETP Lett.\  {\bf 50} (1989) 346
[Pisma Zh.\ Eksp.\ Teor.\ Fiz.\  {\bf 50} (1989) 312];

\bibitem{Bizon:1990sr}
P.~Bizon,
``Colored black holes,''
Phys.\ Rev.\ Lett.\  {\bf 64} (1990) 2844.
\doi{10.1103/PhysRevLett.64.2844}

\bibitem{Meessen:2008kb}
P.~Meessen,
``Supersymmetric coloured/hairy black holes,''
Phys.\ Lett.\ B {\bf 665} (2008) 388.
\doi{10.1016/j.physletb.2008.06.035}.
[\arxiv{0803.0684} [hep-th]].

\bibitem{Meessen:2015nla}
P.~Meessen and T.~Ort\'{\i}n,
``$ \mathcal{N}=2 $ super-EYM coloured black holes from defective Lax matrices,''
JHEP {\bf 1504} (2015) 100.
\doi{10.1007/JHEP04(2015)100}.
[\arxiv{1501.02078} [hep-th]].

\bibitem{Meessen:2015enl}
P.~Meessen, T.~Ort\'{\i}n and P.~Fern\'andez-Ram\'{\i}rez,
``Non-Abelian, supersymmetric black holes and strings in 5 dimensions,''
JHEP {\bf 1603} (2016) 112.
\doi{10.1007/JHEP03(2016)112}.
[\arxiv{1512.07131} [hep-th]].

\bibitem{Ortin:2016bnl}
T.~Ort\'{\i}n and P.~F.~Ram\'{\i}rez,
``A non-Abelian Black Ring,''
Phys.\ Lett.\ B {\bf 760} (2016) 475.
\doi{10.1016/j.physletb.2016.07.018}
[\arxiv{1605.00005} [hep-th]].

\bibitem{Callan:1996dv}
C.~G.~Callan and J.~M.~Maldacena,
``D-brane approach to black hole quantum mechanics,''
Nucl.\ Phys.\ B {\bf 472} (1996) 591.
\doi{10.1016/0550-3213(96)00225-8}
[\hepth{9602043}].

\bibitem{David:2002wn}
J.~R.~David, G.~Mandal and S.~R.~Wadia,
``Microscopic formulation of black holes in string theory,''
Phys.\ Rept.\  {\bf 369} (2002) 549.
\doi{10.1016/S0370-1573(02)00271-5}
[\hepth{0203048}].

\bibitem{Belavin:1975fg}
A.~A.~Belavin, A.~M.~Polyakov, A.~S.~Schwartz and Y.~S.~Tyupkin,
``Pseudoparticle Solutions of the Yang-Mills Equations,''
Phys.\ Lett.\ B {\bf 59} (1975) 85.
\doi{10.1016/0370-2693(75)90163-X}.

\bibitem{Cano:2017sqy}
P.~A.~Cano, T.~Ort\'{\i}n and P.~F.~Ram\'{\i}rez,
``A gravitating Yang-Mills instanton,''
\arxiv{1704.00504} [hep-th].

\bibitem{kn:CMOR}
P.~A.~Cano, P.~Meessen, T.~Ort\'{\i}n and P.~F.~Ram\'{\i}rez,
``$\alpha'$-corrected non-Abelian black holes in string theory'',
in preparation.

\bibitem{Strominger:1990et}
A.~Strominger,
``Heterotic solitons,''
Nucl.\ Phys.\ B {\bf 343} (1990) 167.
Erratum: [Nucl.\ Phys.\ B {\bf 353} (1991) 565].
\doi{10.1016/0550-3213(91)90349-3}, \doi{10.1016/0550-3213(90)90599-9}

\bibitem{Halmagyi:2016pqu}
N.~Halmagyi, D.~Israel and E.~E.~Svanes,
``The Abelian Heterotic Conifold,''
JHEP {\bf 1607} (2016) 029.
\doi{10.1007/JHEP07(2016)029}
[\arxiv{1601.07561} [hep-th]].


\bibitem{Dabholkar:1995ep}
A.~Dabholkar,
``Ten-dimensional heterotic string as a soliton,''
Phys.\ Lett.\ B {\bf 357} (1995) 307.
\doi{10.1016/0370-2693(95)00949-L}
[\hepth{9506160}].

\bibitem{Hull:1995nu}
C.~M.~Hull,
``String-string duality in ten-dimensions,''
Phys.\ Lett.\ B {\bf 357} (1995) 545.
\doi{10.1016/0370-2693(95)01000-G}
[\hepth{9506194}].

\bibitem{Polchinski:1995df}
J.~Polchinski and E.~Witten,
``Evidence for heterotic - type I string duality,''
Nucl.\ Phys.\ B {\bf 460} (1996) 525.
\doi{10.1016/0550-3213(95)00614-1}.
[\hepth{9510169}].

\bibitem{Cano:2016rls}
P.~A.~Cano, T.~Ort\'{\i}n and C.~Santoli,
``Non-Abelian black string solutions of $ \mathcal{N} = (2,0), d = 6$ 
supergravity,''
JHEP {\bf 1612} (2016) 112.
\doi{10.1007/JHEP12(2016)112}
[\arxiv{1607.02595} [hep-th]].

\bibitem{Bueno:2015wva}
P.~Bueno, P.~Meessen, T.~Ort\'{\i}n and P.~F.~Ram\'{\i}rez,
``Resolution of SU(2) monopole singularities by oxidation,''
Phys.\ Lett.\ B {\bf 746} (2015) 109.
\doi{10.1016/j.physletb.2015.04.065}.
[\arxiv{1503.01044} [hep-th]].

\bibitem{Hubscher:2008yz}
M.~H\"ubscher, P.~Meessen, T.~Ort\'{\i}n and S.~Vaul\`a,
``N=2 Einstein-Yang-Mills's BPS solutions,''
JHEP {\bf 0809} (2008) 099.
\doi{10.1088/1126-6708/2008/09/099}.
[\arxiv{0806.1477} [hep-th]].

\bibitem{Bueno:2014mea}
P.~Bueno, P.~Meessen, T.~Ort\'{\i}n and P.~F.~Ram\'{\i}rez,
``$ \mathcal{N}=2 $ Einstein-Yang-Mills' static two-center solutions,''
JHEP {\bf 1412} (2014) 093.
\doi{10.1007/JHEP12(2014)093}
[\arxiv{1410.4160} [hep-th]].

\bibitem{Maldacena:1996ky}
J.~M.~Maldacena,
``Black holes in string theory,'', Ph.D.~Thesis, Princeton University, 1996.
\hepth{9607235}.

\bibitem{Rey:1989xj}
S.~J.~Rey,
``The Confining Phase of Superstrings and Axionic Strings,''
Phys.\ Rev.\ D {\bf 43} (1991) 526.
\doi{10.1103/PhysRevD.43.526}

\bibitem{Tseytlin:1996as}
A.~A.~Tseytlin,
``Extreme dyonic black holes in string theory,''
Mod.\ Phys.\ Lett.\ A {\bf 11} (1996) 689.
\doi{10.1142/S0217732396000709}
[\hepth{9601177}].

\bibitem{Cvetic:1995bj}
M.~Cvetic and A.~A.~Tseytlin,
``Solitonic strings and BPS saturated dyonic black holes,''
Phys.\ Rev.\ D {\bf 53} (1996) 5619.
Erratum: [Phys.\ Rev.\ D {\bf 55} (1997) 3907].
\doi{10.1103/PhysRevD.53.5619}, \doi{10.1103/PhysRevD.55.3907}
[\hepth{9512031}].

\bibitem{Strominger:1996sh}
A.~Strominger and C.~Vafa,
``Microscopic origin of the Bekenstein-Hawking entropy,''
Phys.\ Lett.\ B {\bf 379} (1996) 99.
\doi{10.1016/0370-2693(96)00345-0}
[\hepth{9601029}].

\bibitem{Peet:2000hn}
A.~W.~Peet,
``TASI lectures on black holes in string theory,''
Proceedings of \textsl{Strings, branes and gravity, TASI'99}, 
Boulder, USA, 1999.
J.A.~Harvey, S.~Kachru and E.~Silverstein eds.
World Scientific (2001).
\hepth{0008241}.

\bibitem{Angelantonj:2002ct}
C.~Angelantonj and A.~Sagnotti,
``Open strings,''
Phys.\ Rept.\  {\bf 371} (2002) 1
Erratum: [Phys.\ Rept.\  {\bf 376} (2003) no.6,  407]
\doi{10.1016/S0370-1573(02)00273-9}, \doi{10.1016/S0370-1573(03)00006-1}
[\hepth{0204089}].

\bibitem{Witten:1995gx}
E.~Witten,
``Small instantons in string theory,''
Nucl.\ Phys.\ B {\bf 460} (1996) 541.
\doi{10.1016/0550-3213(95)00625-7}.
[\hepth{9511030}].

\bibitem{Huebscher:2007hj}
M.~H\"ubscher, P.~Meessen, T.~Ort\'{\i}n and S.~Vaul\`a,
``Supersymmetric N=2 Einstein-Yang-Mills monopoles and covariant attractors,''
Phys.\ Rev.\ D {\bf 78} (2008) 065031.
\doi{10.1103/PhysRevD.78.065031}.
[\arxiv{0712.1530} [hep-th]].

\bibitem{Ramirez:2016tqc}
P.~F.~Ram\'{\i}rez,
``Non-Abelian bubbles in microstate geometries,''
JHEP {\bf 1611} (2016) 152.
\doi{10.1007/JHEP11(2016)152}
[\arxiv{1608.01330} [hep-th]].

\bibitem{Bergshoeff:1989de}
E.~A.~Bergshoeff and M.~de Roo,
``The Quartic Effective Action of the Heterotic String and Supersymmetry,''
Nucl.\ Phys.\ B {\bf 328} (1989) 439.
\doi{10.1016/0550-3213(89)90336-2}





\bibitem{Chamseddine:1980cp}
A.~H.~Chamseddine,
``N=4 Supergravity Coupled to N=4 Matter,''
Nucl.\ Phys.\ B {\bf 185} (1981) 403.
\doi{10.1016/0550-3213(81)90326-6}

\bibitem{Bergshoeff:1981um}
E.~Bergshoeff, M.~de Roo, B.~de Wit and P.~van Nieuwenhuizen,
``Ten-Dimensional Maxwell-Einstein Supergravity, Its Currents, 
and the Issue of Its Auxiliary Fields,''
Nucl.\ Phys.\ B {\bf 195} (1982) 97.
\doi{10.1016/0550-3213(82)90050-5}

\bibitem{Chapline:1982ww}
G.~F.~Chapline and N.~S.~Manton,
``Unification of Yang-Mills Theory and Supergravity in Ten-Dimensions,''
Phys.\ Lett.\  {\bf 120B} (1983) 105.
\doi{10.1016/0370-2693(83)90633-0}

\bibitem{Green:1984sg}
M.~B.~Green and J.~H.~Schwarz,
``Anomaly Cancellation in Supersymmetric D=10 Gauge Theory 
and Superstring Theory,''
Phys.\ Lett.\  {\bf 149B} (1984) 117.
\doi{10.1016/0370-2693(84)91565-X}

\bibitem{Bergshoeff:1988nn}
E.~Bergshoeff and M.~de Roo,
``Supersymmetric Chern-simons Terms in Ten-dimensions,''
Phys.\ Lett.\ B {\bf 218} (1989) 210.
\doi{10.1016/0370-2693(89)91420-2}

\bibitem{Bergshoeff:1992cw}
E.~A.~Bergshoeff, R.~Kallosh and T.~Ort\'{\i}n,
``Supersymmetric string waves,''
Phys.\ Rev.\ D {\bf 47} (1993) 5444.
\doi{10.1103/PhysRevD.47.5444}
[\hepth{9212030}].


\end{thebibliography}
\end{document}